# PYSED: A tool for extracting kinetic-energy-weighted phonon dispersion and lifetime from molecular dynamics simulations

Ting Liang,[1,*] Wenwu Jiang,[2,*] Ke Xu,[1] Hekai Bu,[2] Zheyong Fan,[3] Wengen Ouyang,[2,4,†] and Jianbin Xu[1,‡]

[1]*Department of Electronic Engineering and Materials Science and Technology Research Center, The Chinese University of Hong Kong, Shatin, N.T., Hong Kong SAR, 999077, P. R. China*

[2]*Department of Engineering Mechanics, School of Civil Engineering, Wuhan University, Wuhan, Hubei 430072, China*

[3]*College of Physical Science and Technology, Bohai University, Jinzhou 121013, P. R. China*

[4]*State Key Laboratory of Water Resources Engineering and Management, Wuhan University, Wuhan, Hubei 430072, China*

Machine learning potential-driven molecular dynamics (MD) simulations have significantly enhanced the predictive accuracy of thermal transport properties across diverse materials. However, extracting phonon-mode-resolved insights from these simulations remains a critical challenge. Here, we introduce PYSED, a Python-based package built on the spectral energy density (SED) method, designed to efficiently compute kinetic-energy-weighted phonon dispersion and extract phonon lifetime from large-scale MD simulation trajectories. By integrating high-accuracy machine-learned neuroevolution potential (NEP) models, we validate and showcase the effectiveness of the implemented SED method across systems of varying dimensionalities. Specifically, the NEP-driven MD-SED accurately reveals how phonon modes are affected by strain in carbon nanotubes, as well as by interlayer coupling strength and twist angle in two-dimensional molybdenum disulfide. For three-dimensional systems, the SED method effectively establishes the thermal transport regime diagram for metal-organic frameworks, distinguishing between particlelike and wavelike propagation regions. Moreover, using bulk silicon as an example, we show that phonon SED can efficiently capture quantum dynamics based on path-integral trajectories. The PYSED package bridges MD simulations with detailed phonon-mode insights, delivering a robust tool for investigating thermal transport properties with detailed mechanisms across various materials.

## I. INTRODUCTION

With the recent rapid advancements in machine-learned potentials (MLPs), MLP-driven molecular dynamics (MD) has emerged as a popular approach for providing reliable predictions of thermal transport properties across diverse materials [1–7]. This success stems from MLPs' ability to achieve near-quantum-mechanical accuracy in modeling interatomic interactions, while offering computational efficiency that surpasses density functional theory (DFT) calculations by several orders of magnitude [8, 9]. Prominent MD-based methods for thermal conductivity calculations encompass homogeneous non-equilibrium MD (HNEMD) grounded in linear response theory, equilibrium MD (EMD) based on the Green-Kubo formalism, and non-equilibrium MD (NEMD) following Fourier's law of heat conduction [1, 10]. However, unlike anharmonic lattice dynamics (ALD) calculations (primarily applicable at low temperatures where higher-order anharmonic interactions remain relatively weak), which can naturally provide phonon-mode-dependent information, these MD methods offer limited insights into phonon-mode-level properties [11–13].

Spectral energy density (SED) is a technique that directly predicts phonon dispersion and the corresponding lifetime from atomic trajectories obtained in large-scale MD simulations [11–17]. It can be interpreted as the intensity of the kinetic energy distribution of a system over different phonon modes, that is, a dispersion profile weighted by the kinetic energy carried by phonons [11, 12]. Consequently, SED offers direct insights into phonon-mode-level properties from MD simulations, fully capturing the anharmonicity of interatomic interactions and other phonon scattering processes. This capability becomes particularly critical, as MLP-driven MD simulations advance thermal conductivity predictions, while simultaneously creating an urgent demand for tools capable of acquiring phonon-mode-resolved information. Although several tools, such as DYNASOR [18, 19], DYNAPHOPY [20], and the Nanohub web-based platform [21], have indirectly or directly implemented the SED method, efficient and streamlined tools for computing SED from MD trajectories remain scarce.

To contribute to this community, we present PYSED—a Python-based computational package specifically designed to compute kinetic-energy-weighted phonon dispersion and extract phonon lifetime efficiently. This package demonstrates good compatibility with mainstream MD packages, capable of processing trajectory data generated by LAMMPS [22] and GPUMD [23]. By leveraging the high-efficiency neuroevolution potential (NEP), we systematically validate PYSED's capabilities across systems of varying dimensionality, encompassing typical one-dimensional (1D) carbon nanotubes (CNTs), two-dimensional (2D) materials such as graphene, hexagonal boron nitride ($h$-BN), and molybdenum disulfide ($MoS_2$), and three-dimensional (3D) systems including strongly anharmonic metal-organic frameworks (MOFs) and widely utilized bulk silicon.

* These authors contributed equally to this work.
† w.g.ouyang@whu.edu.cn
‡ jbxu@ee.cuhk.edu.hk

Using high-precision trajectories from NEP-MD simulations, the SED method reveals how compressive strain reduces phonon lifetimes in CNTs. For 2D systems, the SED approach accurately reproduces phonon dispersions for graphene and $h$-BN, matching results from lattice dynamics calculations and experimental measurements. We further demonstrate that the SED method adeptly captures the effects of interlayer coupling strength and twist angles on various phonon modes in $MoS_2$. In 3D systems, the SED method efficiently constructs the thermal transport spectrum for MOFs, demonstrating its applicability to a wide range of strongly anharmonic materials. Finally, leveraging quantum trajectories from path integral molecular dynamics (PIMD) simulations, we employ PYSED to investigate the impact of nuclear quantum effects (NQEs) on the phonon properties of silicon, demonstrating that the SED method effectively captures both quantum and classical dynamics, depending on the characteristics of the input trajectories. Through a series of case studies, we showcase the versatility of PYSED across a variety of systems. With further extensions, PYSED can also utilize atomic trajectories obtained from empirical potential-based MD and DFT-driven MD simulations.

## II. METHODOLOGIES AND MODELS

### A. Phonon Spectral Energy Density Technique

In this section, we briefly derive two equivalent expressions for phonon SED: one requiring eigenvectors and another that does not, while summarizing their respective advantages and disadvantages. We further clarify the selection of allowed wave vectors in periodic systems using a Diophantine-equation approach to identify commensurate $\mathbf{q}$ points, similar to the strategy used in the DYNASOR package [18, 19]. The necessity of Lorentzian fitting for extracting phonon lifetimes in the SED method is subsequently elucidated.

#### 1. Derivation of the SED formalism

According to lattice dynamics, the atomic velocities in real space can be projected onto the time-dependent normal mode velocities [24],

$$\dot{q}(\mathbf{q},\nu,t) = \sum_{\alpha,b,l}^{3,n,N_T} \sqrt{\frac{m_b}{N_T}}\, \dot{u}_\alpha(l,b,t) e^*_\alpha(\mathbf{q},\nu,b) \exp(i\mathbf{q}\cdot\mathbf{r}_0(l)). \tag{1}$$

Here, $m_b$ represents the mass of the $b$-th basis atom, $\dot{u}_\alpha(l,b,t)$ is the $\alpha$-th Cartesian direction of the velocity of the $b$-th basis atom in the $l$-th unit cell at time $t$, $e^*_\alpha(\mathbf{q},\nu,b)$ is the complex conjugate of the $\alpha$ component of the eigenvector corresponding to the $b$-th basis atom, while $\mathbf{r}_0(l)$ denotes the equilibrium position of the $l$-th unit cell. The variable $\nu$ represents the phonon branch, and $n$ and $N_T$ are the number of basis atoms and the number of unit cells in the system, respectively.

The total kinetic energy of the system can be written as the sum of the kinetic energies of the individual normal modes. For a specific normal mode characterized by wave vector $\mathbf{q}$ and branch $\nu$, the time-averaged kinetic energy is,

$$\overline{T}(\mathbf{q},\nu) = \frac{1}{2}\left\langle |\dot{q}(\mathbf{q},\nu,t)|^2 \right\rangle \tag{2}$$

$$= \frac{1}{2\tau_0}\int_0^{\tau_0} |\dot{q}(\mathbf{q},\nu,t)|^2\, dt, \tag{3}$$

where $\langle\cdot\rangle$ denotes the time average and $\tau_0$ is the total MD simulation time. To derive the kinetic energy in the frequency domain, one can perform a Fourier transform on the time-dependent normal mode velocity $\dot{q}(\mathbf{q},\nu,t)$,

$$\tilde{q}(\mathbf{q},\nu,\omega) = \frac{1}{\sqrt{2\pi}}\int_0^{\tau_0} \dot{q}(\mathbf{q},\nu,t)\exp(-i\omega t)\, dt. \tag{4}$$

Then, Parseval's theorem is applied to relate the time-averaged kinetic energy to the energy distribution in the frequency domain,

$$\int_0^{\tau_0} |\dot{q}(\mathbf{q},\nu,t)|^2 dt = \int_{-\infty}^{\infty} |\tilde{q}(\mathbf{q},\nu,\omega)|^2 d\omega. \tag{5}$$

Substituting Eq. (4) and Eq. (5) into Eq. (2) gives

$$\overline{T}(\mathbf{q},\nu) = \frac{1}{4\pi\tau_0}\int_{-\infty}^{\infty}\left|\int_0^{\tau_0} \dot{q}(\mathbf{q},\nu,t)\exp(-i\omega t)dt\right|^2 d\omega. \tag{6}$$

To calculate the total SED $\Phi(\mathbf{q},\omega)$ for a given $\mathbf{q}$ point, one can substitute Eq. (1) into Eq. (6) and sum over the contributions of all phonon branches [13, 25],

$$\begin{aligned}\Phi(\mathbf{q},\omega) &= \sum_{\nu}^{3n} \overline{T}(\mathbf{q},\nu) \\ &= \frac{1}{4\pi\tau_0 N_T}\sum_{\nu}^{3n}\Bigg|\sum_{\alpha}^{3}\sum_{b}^{n}\int_0^{\tau_0}\sum_{l}^{N_T}\sqrt{m_b} \\ &\quad\times \dot{u}_\alpha(l,b,t)e^*_\alpha(\mathbf{q},\nu,b)\exp\big(i\mathbf{q}\cdot\mathbf{r}_0(l) - i\omega t\big)\, dt\Bigg|^2.\end{aligned} \tag{7}$$

Currently, phonon SED has two expressions, with Eq. (7) derived from the frequency-domain normal mode decomposition (NMD) as the first version [12, 13, 26], which includes the eigenvectors. Compared to the first version, the second version, derived by Thomas et al. [11], calculates the total SED by summing over the basis atoms in the unit cell without determining the SED of each phonon branch,

$$\Phi'(\mathbf{q},\omega) = \frac{1}{4\pi\tau_0 N_T}\sum_{\alpha}^{3}\sum_{b}^{n} m_b \left| \int_0^{\tau_0} \sum_{l}^{N_T} \dot{u}_\alpha(l,b,t) \times \exp\left(i\mathbf{q}\cdot\mathbf{r}_0(l) - i\omega t\right) dt \right|^2 . \quad (8)$$

Utilizing the orthonormal condition of the eigenvectors [11, 13, 24],

$$\sum_{\nu}^{3n} e_\alpha(\mathbf{q},\nu,b) e_\beta^*(\mathbf{q},\nu,b) = \delta_{\alpha\beta}, \quad (9)$$

where $\delta_{\alpha\beta}$ is the Kronecker delta function, Eq. (7) simplifies to Eq. (8), thereby eliminating the eigenvectors. It is worth noting that $\Phi'(\mathbf{q},\omega)$ calculates the average kinetic energy per unit cell as a function of wave vector $\mathbf{q}$ and frequency, rather than the contribution from each individual atom, so only the entire integral term is squared.

There has been an ongoing debate on whether the two versions of SED can produce the same results [12]. By employing Eq. (9), Feng et al. [13] analytically proved that the second version $\Phi'(\mathbf{q},\omega)$, which does not depend on eigenvectors, can yield the same result as the first version $\Phi(\mathbf{q},\omega)$ within numerical precision. However, if the frequencies of two phonon modes are so close that their SED peaks cannot be distinguished in the entire SED spectrum, applying the eigenvectors can help to separate them into two distinct plots. Fitting $\Phi'(\mathbf{q},\omega)$ at higher temperatures becomes challenging, as phonon linewidths become wider and are comparable to the interval between mode frequencies.

Given the advancements in high-accuracy MLPs alongside robust thermal conductivity calculation techniques such as HNEMD, EMD, and NEMD [1, 10], we do not recommend using the SED method for thermal conductivity prediction. Therefore, relying on SED for highly accurate phonon lifetime evaluations is not a strict requirement. Instead, we recommend using the second version $\Phi'(\mathbf{q},\omega)$ to evaluate the phonon mode properties, as it does not require the specific calculation of eigenvectors. Especially for complex systems, the computational cost of eigenvectors is very high. Consequently, the PYSED package currently integrates only the second version of the SED formulation, $\Phi'(\mathbf{q},\omega)$.

Additionally, by analyzing the autocorrelation function of phonon energy, one can describe the phonon energy in the time domain using the NMD method [14–17], while $\Phi$ and $\Phi'$ represent the phonon energy in the frequency domain. The time-domain and frequency-domain approaches are mathematically equivalent, as the Wiener-Khinchin theorem dictates [27–29]. The frequency-domain method offers the advantage of enabling phonon lifetime and frequency predictions by fitting simpler functions, in contrast to the time-domain method.

### *2. Allowed Wave Vectors*

Based on the Born-Von Kármán boundary conditions, the allowed quantized wave vectors in periodic systems are discretized and for 1D can be defined by [12, 24],

$$q_\alpha = \frac{2\pi}{N_\alpha a_\alpha} n_\alpha, \quad (10)$$

where $a_\alpha$ is the lattice constant, $N_\alpha$ is the total number of unit cells in the $\alpha$ direction, and $n_\alpha$ is an integer ranging from $-N_\alpha/2 < n_\alpha \le N_\alpha/2$. Due to the time-reversal symmetry present in most crystal systems, many practical implementations only take $n_\alpha \ge 0$ and reduce the set of wave vectors to avoid redundancy.

In 3D systems, the discretized wave vector $\mathbf{q}$ must satisfy the condition of being an allowed point in the reciprocal space of the supercell and is typically restricted to the first Brillouin zone. In PYSED, the supercell is constructed via

$$\mathbf{S} = \mathbf{P}\cdot\mathbf{p} \quad (11)$$

where $\mathbf{p}$ is a $3\times 3$ matrix whose rows are the primitive lattice vectors, and $\mathbf{S}$ is a $3\times 3$ matrix whose rows are the supercell lattice vectors. The integer matrix $\mathbf{P}$ specifies how many times the primitive cell is replicated along each direction. A wave vector $\mathbf{q}$ is considered commensurate with the supercell if

$$\mathbf{q}_{\text{red}}\,\mathbf{P}^{\top} \in \mathbb{Z}^3, \quad (12)$$

where $\mathbf{q}_{\text{red}}$ is the wave vector in "reduced coordinates" relative to the primitive reciprocal lattice. The condition "$\in \mathbb{Z}^3$" ensures that each component is an integer, matching the discrete reciprocal grid of the supercell.

To locate the commensurate wave vectors along a path from $\mathbf{q}_{\text{start}}$ to $\mathbf{q}_{\text{end}}$ (between two high-symmetry points and given in reduced coordinates), PYSED performs the following steps:

1. Parametric line search and Diophantine equation. The code considers the line

$$\mathbf{q}(f) = \mathbf{q}_{\text{start}} + f\left(\mathbf{q}_{\text{end}} - \mathbf{q}_{\text{start}}\right), \quad (13)$$

and identifies all fractional $f$ in $[0,1]$ for which

$$\mathbf{q}(f)\,\mathbf{P}^{\top} \in \mathbb{Z}^3. \quad (14)$$

Equivalently, by writing

$$\mathbf{A} = \mathbf{q}_{\text{start}}\,\mathbf{P}^{\top}, \quad \mathbf{B} = \left(\mathbf{q}_{\text{end}} - \mathbf{q}_{\text{start}}\right)\mathbf{P}^{\top}, \quad (15)$$

the condition $\mathbf{A} + f\,\mathbf{B} \in \mathbb{Z}^3$ constitutes a linear Diophantine-like equation in fractional form. Numerically, PYSED solves for all valid $f$ that yield integer components.

2. Collect commensurate $\mathbf{q}$ points.
   For each valid $f$, the wave vector $\mathbf{q}(f)$ is mapped into Cartesian space,

$$\mathbf{q}_{\mathrm{cart}}(f) \;=\; \mathbf{q}(f)\left[2\pi\,(\mathbf{p}^{-1})^{\mathsf{T}}\right], \tag{16}$$

   and recorded. Repeating this for multiple segments (e.g., $\Gamma \to X$, $X \to M$, etc.) yields a piecewise path through the first Brillouin zone.

Thus, analogous to the algorithm implemented in the DYNASOR package [18, 19], PYSED employs a Diophantine-equation approach in fractional form to determine the specific values of $f$ for which $\mathbf{q}(f)$ is an allowed wave vector on the supercell lattice. This method supports both orthogonal and non-orthogonal lattices and provides a discrete set of $\mathbf{q}$ points suitable for subsequent SED calculations. Without loss of generality, the larger the supercell matrix $\mathbf{P}$, the greater the total number of allowed discrete wave vectors $\mathbf{q}$, but the memory consumption also increases.

#### 3. *Phonon Lifetime Extraction – Lorentzian Function Fitting*

The phonon lifetime can be obtained by fitting the SED curve by the Lorentzian function [11],

$$\Phi(\mathbf{q},\omega), \Phi'(\mathbf{q},\omega) = \frac{I}{1+\left[(\omega-\omega_c)/\gamma\right]^2}, \tag{17}$$

where $I$ is the peak magnitude, $\omega_c$ is the frequency at the peak center, and $\gamma$ is the half-width at half-maximum (HWHM). Finally, the $\tau$ at each wave vector $\mathbf{q}$ and frequency $\omega$ is defined as,

$$\tau(\mathbf{q},\omega) = \frac{1}{2\gamma}. \tag{18}$$

### B. NEP Machine-Learned Potential Models

#### 1. *NEP framework*

To construct accurate potential models that drive MD simulations and generate near-first-principles accuracy trajectories, we employ the NEP framework [30]. Based on a feedforward neural network, the NEP method utilizes a single hidden layer with $N_{\mathrm{neu}}$ neurons to represent the site energy $U_i$ of a system containing $N$ atoms, which can be explicitly expressed as:

$$U_i = \sum_{\mu=1}^{N_{\mathrm{neu}}} w_{\mu}^{(1)} \tanh\left(\sum_{\nu=1}^{N_{\mathrm{des}}} w_{\mu\nu}^{(0)} q_{\nu}^{i} - b_{\mu}^{(0)}\right) - b^{(1)}, \tag{19}$$

where $\tanh(x)$ is the activation function, $\mathbf{w}^{(0)}$ are the weight parameters mapping the input layer (with dimension $N_{\mathrm{des}}$) to the hidden layer (with dimension $N_{\mathrm{neu}}$), and $\mathbf{w}^{(1)}$ links the hidden layer to the output layer (the site energy). The bias parameters $\mathbf{b}^{(0)}$ are associated with the hidden layer, and $b^{(1)}$ is the bias term applied to the output layer. All of these parameters, weights, and biases are subject to optimization during the training process.

The input layer corresponds to the descriptor vector $\mathbf{q}_i$ (of dimension $N_{\mathrm{des}}$) for a given atom $i$, with components denoted by $q_{\nu}^{i}$ as defined in Eq. (19) [31]. Each descriptor $q_{\nu}^{i}$ depends on the atomic positions while satisfying invariance under translation, rotation, inversion, and permutation of atoms of the same species. Similar to the symmetry functions in the Behler-Parrinello approach [32], the descriptor $q_{\nu}^{i}$ is constructed from a set of radial and angular components. The NEP method has undergone several improvements, leading to the development of the NEP3 [31] and NEP4 [33] versions. Readers interested in more detailed information on the NEP method and its descriptors are encouraged to consult recent reviews on its applications across various fields [1, 34].

#### 2. *NEP for CNT*

For CNT systems, the general-purpose NEP model for carbon-based materials developed by Fan *et al.* [35] was employed. This model was trained on a comprehensive dataset originally curated for the development of the Gaussian approximation potential [36]. The dataset was generated via DFT calculations with the optB88-vdW dispersion-corrected exchange-correlation functional [37–39]. This NEP model exhibits high fidelity in capturing a broad spectrum of carbon structures, including amorphous carbon, graphene, CNT, and various carbon-based nanostructures. A detailed description of the model's training process, along with extensive benchmarking results, is provided in Refs. [35, 36].

#### 3. *NEP for $MoS_2$*

For $MoS_2$ systems, we use a newly developed hybrid computational framework [40] that integrates a NEP model for intralayer interactions with a registry-dependent interlayer potential (ILP) [41–44] to account for anisotropic interlayer van der Waals interactions. The ILP has been extensively applied in layered van der Waals systems and is well-suited for accurately characterizing interlayer interaction forces in multilayer $MoS_2$ [45–50]. The NEP model was trained exclusively on monolayer configurations. To construct the dataset, we first generated 50 distinct structures by introducing random lattice deformations (ranging from -3% to 3%) and atomic displacements (within 0.1 Å) based on the optimized monolayer $MoS_2$ structure. To further account for thermal fluctuations, we performed MD simulations using empirical potentials [51] under the NVT ensemble, sampling configurations at 300 K, 600 K, and 900 K. At vari-ous temperatures, 200 distinct structures were collected, forming a final dataset comprising 250 structures, all of which were exclusively used as the training set.

To obtain the energy, forces, and stress components required for training, single-point calculations were performed using the Perdew–Burke–Ernzerhof (PBE) exchange-correlation functional [52] within the projector-augmented wave framework, as implemented in the Vienna Ab initio Simulation Package [53, 54] . For the monolayer $MoS_2$ configurations, the simulation box size along the out-of-plane ($z$) direction was set to 50 Å to eliminate spurious interactions between periodic images. The self-consistent electronic energy convergence criterion was set to $10^{-7}$ eV, with a plane-wave energy cutoff of 850 eV. The Brillouin zone was sampled using a Γ-centered $k$-point mesh with a spacing of 0.15/Å. During NEP model training, the cutoff radii for both radial and angular descriptors were set to 5 Å. The final model achieved root-mean-square errors of 0.20 meV/atom for total energy, 20.31 meV/Å for forces, and 4.8 MPa for stress (see Figure S1 in the Supplementary Material for the training parity plot).

#### *4. NEP for graphene and h-BN*

For graphene and $h$-BN systems, the NEP model from Ref. [55] is employed, which was trained on a dataset containing 4099 atomic configurations. This dataset is labeled by PBE functional [52] calculations with state-of-the-art many-body dispersion corrections [56, 57]. The NEP model demonstrates the capability to describe both monolayer and bulk phases of graphene and $h$-BN, yielding phonon dispersions that closely align with experimental measurements [55]. Accordingly, the same NEP model [55] is utilized for both bulk graphite and $h$-BN. Detailed descriptions of the model's training protocol and comprehensive benchmarking are provided in Ref. [55].

#### *5. NEP for MOFs*

For MOFs, we take MOF-5 and HKUST-1 as representative soft porous crystals (with primitive cells containing a large number of atoms) and calculate their phonon dispersion and lifetimes at different temperatures (300, 500 K) to validate the applicability of the SED method. Two NEP models, developed by Ying *et al.* [7], at the PBE level—each tailored to one of the MOFs—are employed to perform large-scale MD simulations. Considering long-range dispersion interactions is crucial for accurately simulating the thermodynamic behavior of MOF crystals [58]. In the MD simulations, we integrate the original NEP models with dispersion interactions by employing the D3 method [59] in conjunction with the Becke-Johnson damping function [60, 61]. Specifically, the cutoff radius for the D3 potential and the calculation of coordination numbers are set to 12 and 6 Å, respectively, to balance accuracy and computational efficiency [59].

#### *6. NEP for silicon*

To investigate the effects of classical MD trajectories and PIMD trajectories on the calculated phonon frequencies and lifetimes of silicon using the SED method at varying temperatures, the NEP model developed by Dong *et al.* [1] is employed. This model, refined through two rounds of active-learning iterations based on MD, has been demonstrated to robustly and efficiently simulate crystalline silicon across a temperature range of 100–1000 K. The quantum trajectories are sampled via thermostatted ring-polymer molecular dynamics (TRPMD) simulations [62].

### C. NEP-SED simulation details

In this section, the supercell configurations and MD simulation protocols for the various studied systems are presented. For the CNT system, we used a $1 \times 1 \times 160$ supercell containing 17,920 atoms for MD simulations. For monolayer graphene and $h$-BN, MD simulations were performed using an $80 \times 80 \times 1$ triclinic supercell containing 12,800 atoms. In contrast, bulk graphite and $h$-BN were simulated using a $15 \times 15 \times 50$ supercell with 45,000 atoms. For the $MoS_2$ with different interlayer coupling strengths, we constructed a large supercell of $15\times15\times50$ with 67,500 atoms, based on an AA'-stacked unit cell (6 atoms). Moreover, supercells of $3\times3\times50$ (containing 99,900 atoms) and $6 \times 6 \times 50$ (containing 75,600 atoms) were employed for twisted $MoS_2$ with twist angles of 9.43° and 21.8°, respectively. The method for constructing the periodically twisted $MoS_2$ system can be found in Ref. [47]. For MOFs, conventional orthogonal cells were adopted. Specifically, MOF-5 was extended to $50 \times 2 \times 2$ supercells (84,800 atoms), while HKUST-1 utilized $40 \times 2 \times 2$ supercells (99,840 atoms). We used a triclinic unit cell containing two atoms for silicon and applied a $30 \times 30 \times 30$ supercell, which includes 54,000 atoms in a simulation model.

In the SED method, employing larger supercells increases the number of allowed discrete wave vectors $\mathbf{q}$, resulting in a finer sampling of phonon dispersion. Therefore, we maximized supercell sizes across all systems within memory constraints to ensure sufficient sampling of commensurate $\mathbf{q}$ points for high-resolution $\Phi'(\mathbf{q}, \omega)$ maps.

During the MD simulations, except for bulk silicon, all systems were first equilibrated at the target temperature under the NpT ensemble (with zero target pressure) and then subjected to a production stage in the NVE ensemble. During this period, atomic velocities and positions were collected every $n$ simulation steps ($\Delta n$). According to the Shannon sampling theorem, the maximum fre-quency resolvable by SED is $\mathrm{Freq}_{\mathrm{max}} \approx 1/(2 \cdot \mathrm{d}t \cdot \Delta n)$, where $\mathrm{d}t$ is the timestep. A longer production time allows for more frames of atomic positions and velocities to be collected, which in turn increases the resolution of the SED maps. However, this comes with the trade-off of higher memory consumption, and based on our experience, it is advisable to keep the total number of trajectory frames between 20,000 and 50,000. For bulk silicon systems, the lattice constant was fixed to ensure that the NQEs observed in the SED method arise from dynamics rather than from thermal expansion. Consequently, classical trajectories for the silicon system were equilibrated in the NVT ensemble before transitioning to NVE for production, while quantum trajectories utilized PIMD for equilibration followed by TRPMD for sampling. In PIMD simulation, we use a ring polymer comprising 64 beads, which is sufficient for the silicon system to achieve convergence even at the lowest temperature considered here. The quantum trajectories were produced via the high-efficiency PIMD algorithm as implemented in the GPUMD package [58].

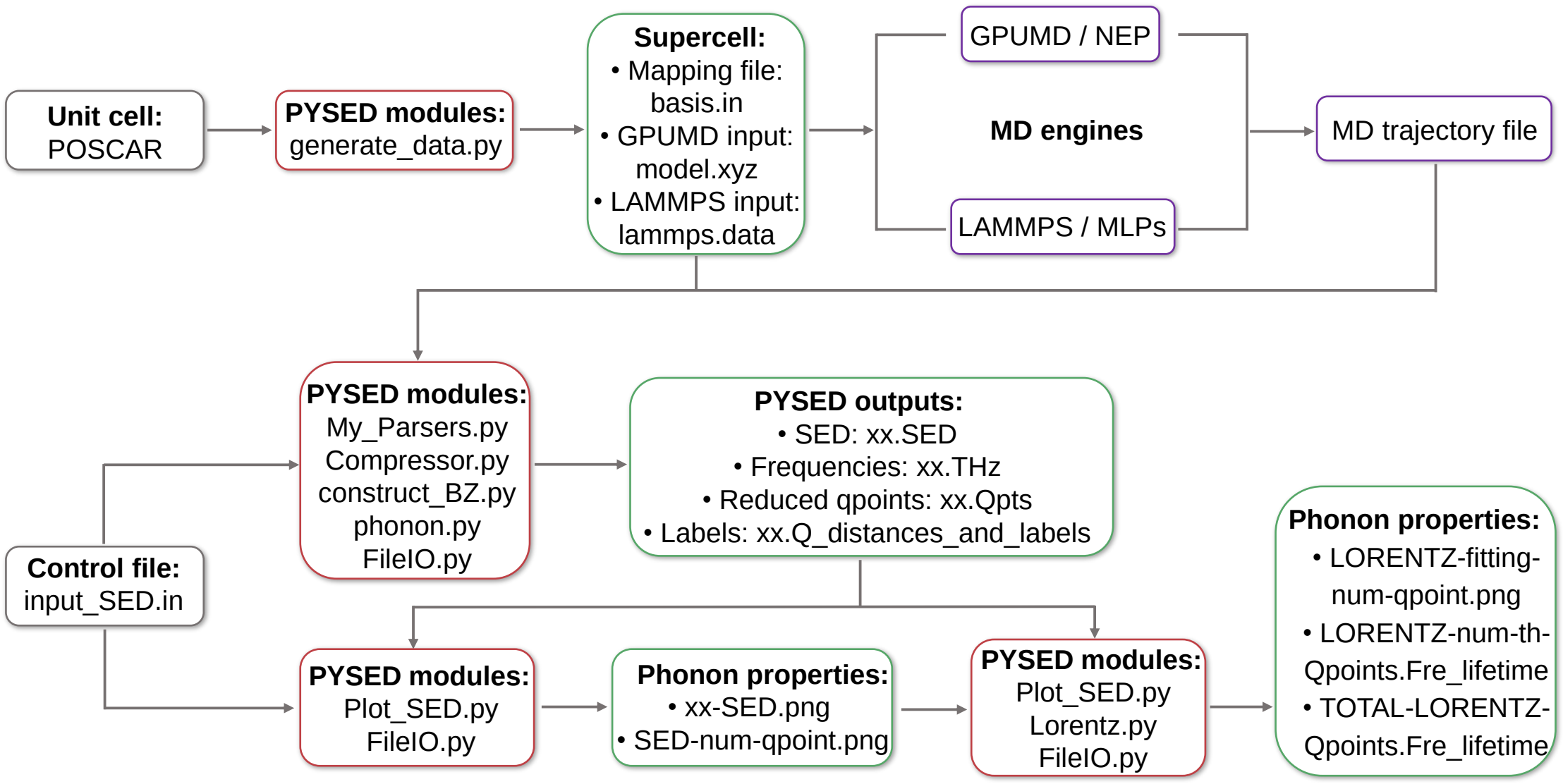

FIG. 1. **Internal workflow of PYSED.** The black rounded boxes denote user inputs, the red rounded boxes represent PYSED modules, the green rounded boxes indicate PYSED outputs, and the purple rounded boxes correspond to external MD engines and trajectories. The "xx" placeholder in output files indicates that their names can be defined in the `input_SED.in` file, which itself may also be customized by the user.

## III. SOFTWARE OVERVIEW

The internal workflow of PYSED is illustrated in Figure 1. Starting with either a primitive or conventional cell as input, the `generate_data.py` module provided by PYSED generates the supercell model required for MD simulations and the mapping file `basis.in`. Subsequently, MLP-driven MD simulations are conducted using the GPUMD [23] and LAMMPS [22] engines, with trajectories produced accordingly. At this stage, simulations can also be performed using empirical potentials.

Next, the user is required to supply PYSED with a control file that includes MD simulation setup, structural information, **q**-point paths, and other necessary parameters. This file is self-explanatory and well-organized. By utilizing PYSED's internal modules and reading `basis.in`, `input_SED.in`, along with the trajectory file, the SED is computed and saved to several local files.

Finally, PYSED's internal modules can be used to visualize the SED map and perform Lorentzian fitting to extract phonon lifetimes. Notably, PYSED leverages functions provided by the SCIPY package [63] to automatically identify and fit the SED peaks. To achieve this, the user only needs to specify three key parameters in the control file `input_SED.in`: the minimum peak height of SED (`peak_height`), the peak prominence indicating how distinctly a peak stands out from the surrounding baseline (`peak_prominence`), and an initial guess for the HWHM used in the Lorentzian fitting (`initial_guess_hwhm`). Extensive testing has led to a reasonably good default value for the initial HWHM in PYSED. However, users are still encouraged to adjust this parameter based on the Lorentzian fitting plots generated for various systems. Furthermore, PYSED is capable of performing SED peak fitting for a single **q** point as well as for all **q** points simultaneously, depending on the user's input.

## IV. APPLICATIONS

In this section, we showcase the versatile applications of PYSED for materials ranging from 1D to 3D, and illus-trate the rich phonon-mode-level information exhibited in the SED of different systems. In all cases presented here, MD simulations are driven by the NEP model using the GPUMD package [23], and only the second version $\Phi'(\mathbf{q}, \omega)$ is employed.

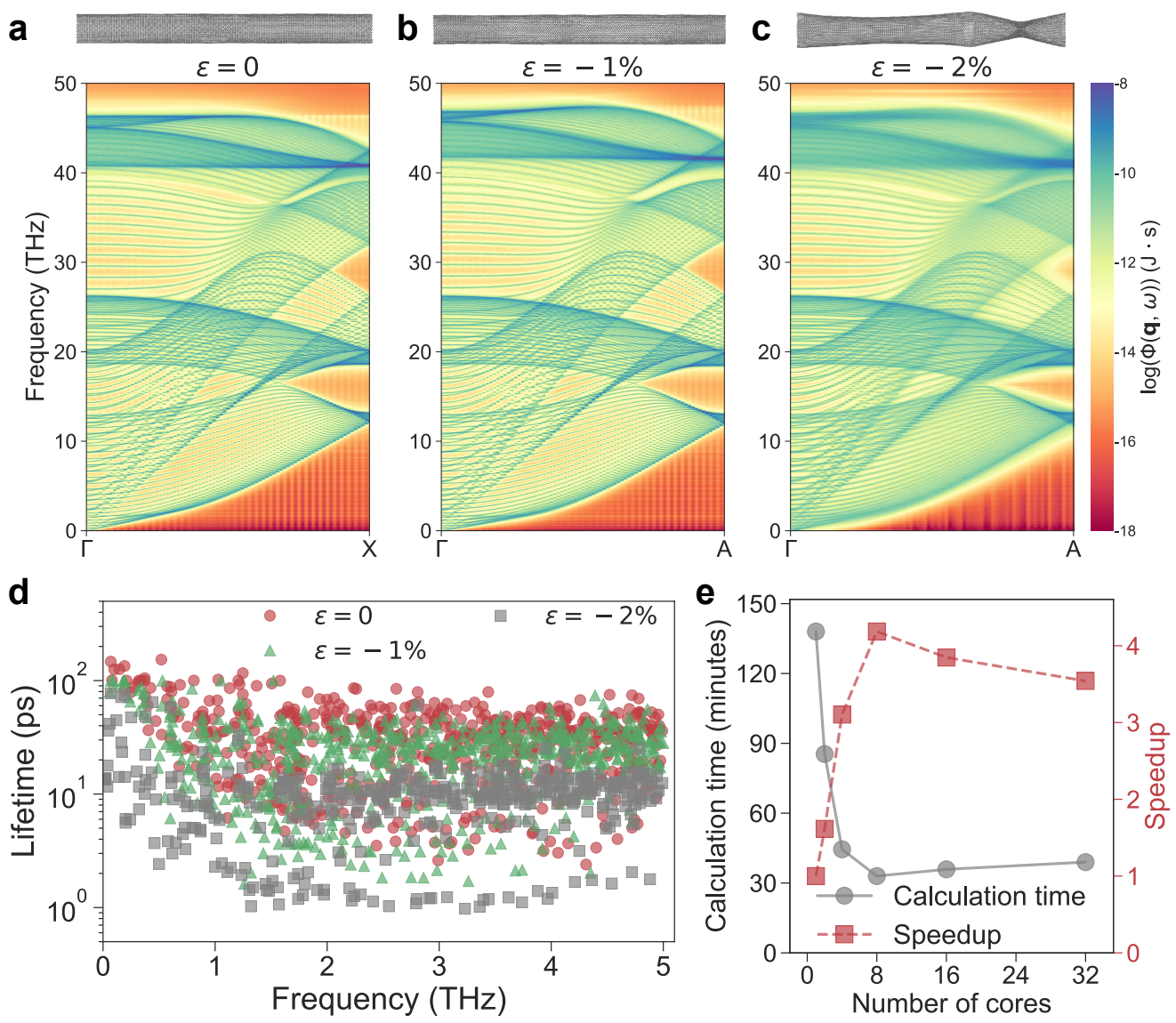

FIG. 2. **a**–**c** Phonon SED and **d** the corresponding phonon lifetimes for CNTs under different compressive strains at 300 K. The top of panels **a**–**c** shows atomic snapshots taken at different compressive strain levels. **e** The calculation time and speedup using varying numbers of cores in PYSED package for calculating the SED of CNTs.

### A. Carbon nanotube

We first investigate the SED of a single-walled CNT with chirality $(28, 28)$ (diameter: 3.74 nm) under compressive strains of 0%, −1%, and −2%. By employing PYSED's integrated plotting module (see Figure 1), the SED of the CNTs reveals clear phonon dispersion (Figure 2a–c). This is because if distinct maxima are present in the SED, clear discrete points would appear on the $(\mathbf{q}, \omega)$ plane, subsequently forming dispersion curves [4, 64]. The comparison of phonon dispersions obtained from SED and lattice dynamics driven by different supercells is presented in Figure S2. When the supercell used in lattice dynamics reaches or exceeds $1 \times 1 \times 4$, the resulting dispersions agree well with the SED results, underscoring the importance of careful supercell-size validation in lattice dynamics calculations (see Figure S2).

As shown in Figure 2a–c, with increasing compressive strain, the discrete points in the CNTs' SED become progressively less distinct, implying enhanced peak broadening (confirmed by Figure S3). Correspondingly, the phonon lifetimes derived from Lorentzian fitting diminish with increasing compressive strain (Figure 2d). Owing to geometric instability, CNTs experience local buckling (Figure 2a–c) when compressed, leading to enhanced phonon scattering [65–67] and a subsequent reduction in phonon lifetimes. This characteristic is accurately captured by the NEP-driven SED. Notably, the PYSED package can automatically and precisely identify the peak positions of the SED at different $\mathbf{q}$ points (Figure S3) and perform Lorentzian fitting (using Eq. (17)) on them without the need for manual intervention.

The PYSED package employs a multi-process parallelization scheme to distribute the computation of SED across distinct wave vectors $\mathbf{q}$. As shown in Figure 2e, we evaluate the computational time and achievable speedup when calculating the SED of CNTs using different numbers of CPU cores. The parallel performance test utilizes an MD trajectory with 62,500 frames, totaling approximately 79 GB in size, and involves computing 80 allowed $\mathbf{q}$ points. Using a single core of the Intel Xeon Gold 6420 CPU, the full computation takes nearly 138 minutes, whereas with 8 cores, the time is reduced to about 33 minutes, achieving a speedup exceeding 4 times and significantly improving computational efficiency (Figure 2e).

### B. Graphene and *h*-BN

We now extend the application of PYSED to 2D systems, with an initial focus on monolayer graphene and *h*-BN. As illustrated in Figure 3a, the phonon spectra along the Γ–M–K–Γ path obtained via the SED nearly coincide with the results from NEP-driven lattice dynamics for both graphene and *h*-BN, thereby confirming the robust applicability of the SED approach. More specifically, the SED-derived phonon frequencies at 300 K exhibit subtle yet systematic softening in the high-frequency regime compared to the results from lattice dynamics calculations at 0 K. This systematic deviation stems from the fundamental principle of the SED approach, which fully incorporates anharmonic effects and enables the exploration of temperature-dependent phonon dispersion.

We further investigate the applicability of PYSED to graphite and bulk *h*-BN systems, which necessitates the explicit consideration of long-range dispersion interactions. As demonstrated in Figure 3b, the SED-fitted phonon dispersions along the Γ–A path for both graphite and bulk *h*-BN show basic agreement with experimental measurements [68, 69], owing to the near-quantum-mechanical fidelity of the atomic trajectories generated by the NEP model [55]. To ensure that phonon SED results are comparable to experimental observations, we recommend that users employ an NpT ensemble for thorough stress relaxation of the system before extracting trajectories from MD simulations.

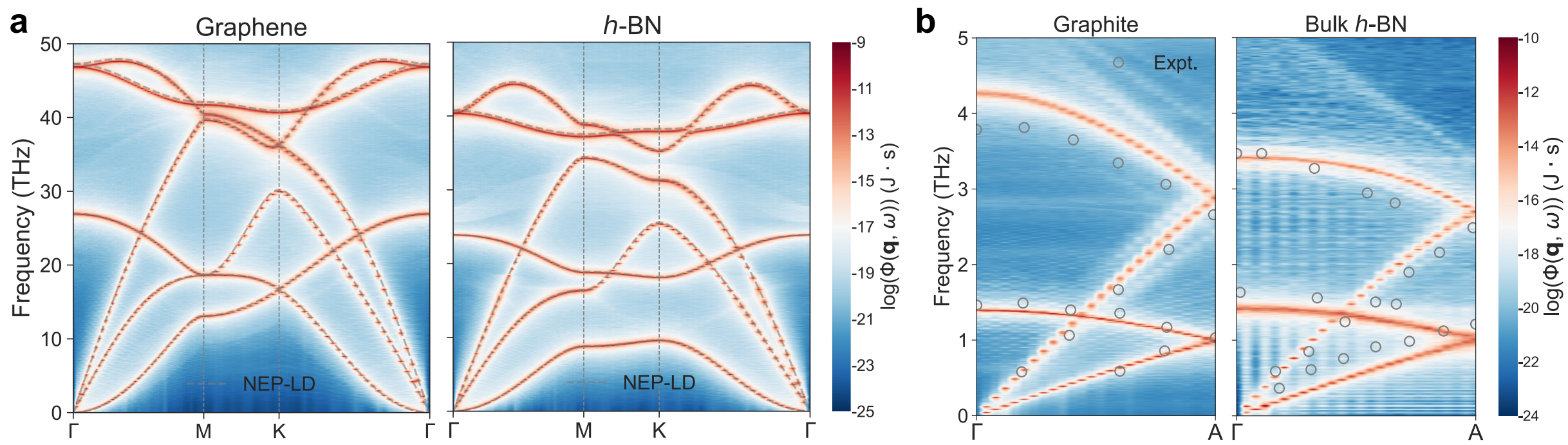

FIG. 3. **a** Comparison of the phonon SED of monolayer graphene and $h$-BN with lattice dynamics calculations. The dashed gray line overlaying the SED map indicates the NEP-driven lattice dynamics calculation results and is labeled "NEP-LD". **b** Comparison of the phonon SED of graphite and bulk $h$-BN with experimental measurements along the Γ to A direction. The experimental data for graphite and bulk $h$-BN are extracted from Refs. [68] and [69], respectively.

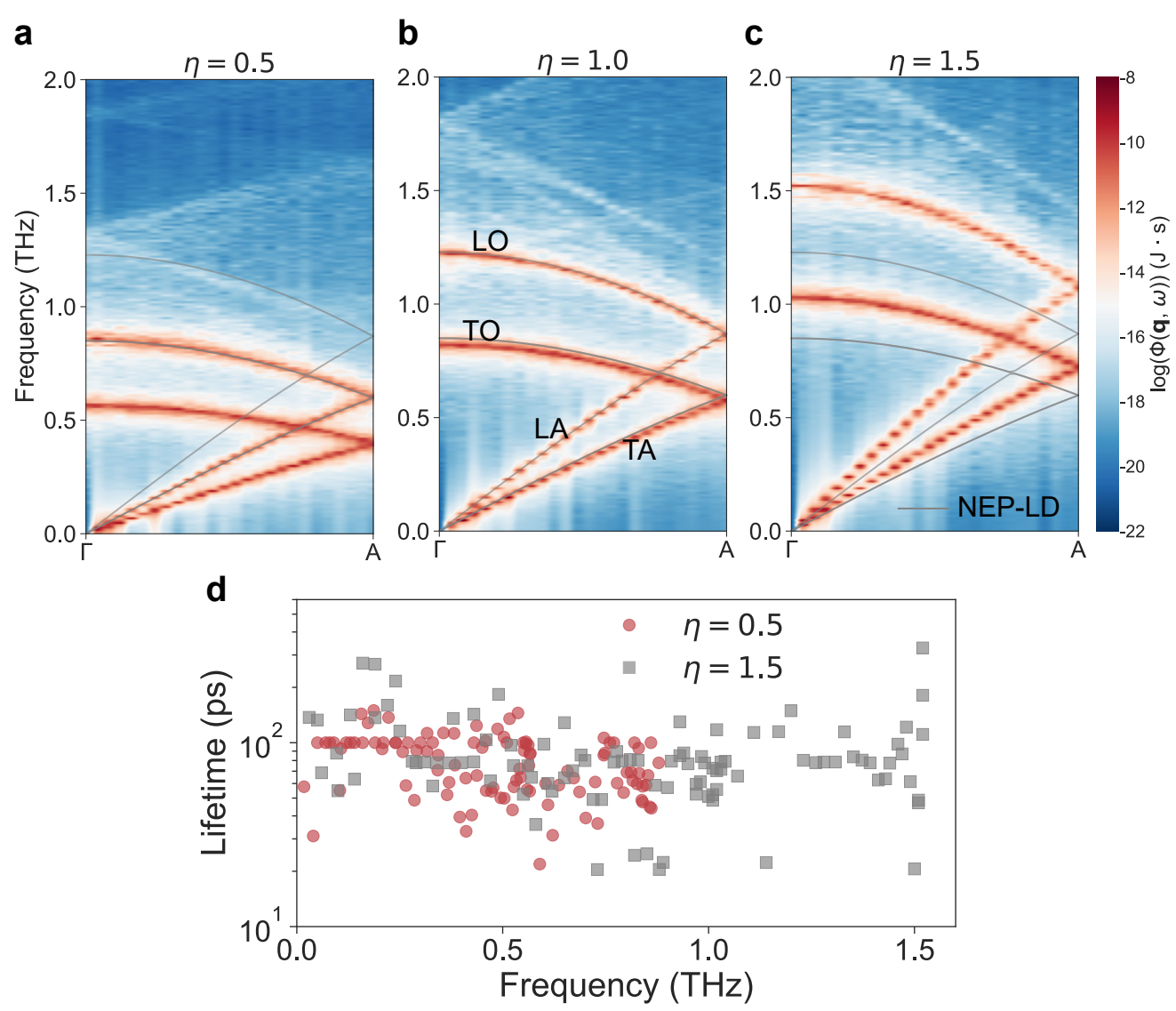

FIG. 4. The effect of different interlayer coupling strengths in bulk AA'-stacked $MoS_2$ on phonon SED and lifetimes along the Γ to A direction around 300 K. Panels **a**–**c** correspond to coupling strengths of 0.5, 1.0, and 1.5, respectively, while panel **d** presents the phonon lifetimes within the 0–1.6 THz range. The gray solid line shows the results of lattice dynamics calculations. The abbreviations TA and TO denote the transverse acoustic and optical phonon branches, respectively, while LA and LO correspond to the longitudinal acoustic and optical phonon branches, respectively.

### C. $MoS_2$

#### 1. *Interlayer coupling effect*

For the bulk $MoS_2$ system, we employ the ILP model (see section II B 3) to account for its anisotropic interlayer van der Waals interactions, where the coupling strength can be tuned by $\eta$. The explicit functional form of the ILP interlayer coupling can be referenced in Ref. [45, 46, 48]. Here, we investigate three $MoS_2$ systems (non-twisted) with distinct interlayer coupling strengths: $\eta =$ 0.5, 1.0, and 1.5, where larger $\eta$ values correspond to stronger interlayer interactions.

At 300 K, both acoustic and optical phonon branches exhibit pronounced collapse (softening) as the coupling strength decreases (Figure 4a–c), corresponding to reduced group velocities [55]. Figure 4d displays the SED-derived phonon lifetimes, where the $\eta = 0.5$ system demonstrates substantially shorter lifetimes compared to $\eta = 1.5$. This disparity arises from weakened interlayer interactions that deteriorate out-of-plane phonon energy transmission. This case study demonstrates that the out-of-plane thermal transport properties in $MoS_2$ systems can be strategically engineered through interlayer coupling modulation, while the phonon SED provides a precise characterization of the underlying mechanisms.

#### 2. *Twisted angle effect*

The effect of twist angles on thermal transport in 2D materials has become a highly active research topic, with extensive studies revealing that the conductance and thermal conductivity of twisted $MoS_2$ systems decrease with increasing twist angles (for $\theta$ roughly less than 30°) [6, 47, 48, 70–72]. Here, we employ the SED method to investigate the underlying mechanisms of this phenomenon. In Figure 5a–c, the out-of-plane LA phonon mode in $MoS_2$ experiences only marginal changes as the twist angle increases. In contrast, the TA modes undergo significant softening and eventual collapse with increasing twist angle, resulting in nearly vanishing group velocities [6, 48, 70]. This physical picture can be attributed to a significant reduction in shear resistance, driven by the twist angle-induced commensurate-to-incommensurate transition, which reduces the energy barriers required for relative shearing between adjacent layers [6, 70].

A further important observation (Figure 5d) is that the phonon lifetimes of TA modes exhibit a nearly order-of-magnitude reduction at $\theta = 9.43°$, yet show unexpected enhancement at $\theta = 21.8°$ compared to the pristine configuration ($\theta = 0°$). The reduced lifetimes originate from symmetry-breaking-induced phonon scattering, evidenced by the substantial spectral broadening of TA modes across multiple **q** points at $\theta = 9.43°$ (see Figure S4). The enhanced lifetimes at $\theta = 21.8°$ arise from extreme spectral narrowing (Figure S4), wherein the TA mode collapse induces a significant compression of the SED peak width near the frequency origin ($\omega \approx 0$). Despite the observed enhancement of phonon lifetimes at large twist angles, the near-zero group velocities of collapsed TA modes drastically suppress the estimated mean free paths. This suppression directly accounts for the severe deterioration of out-of-plane thermal transport as the twist angle increases.

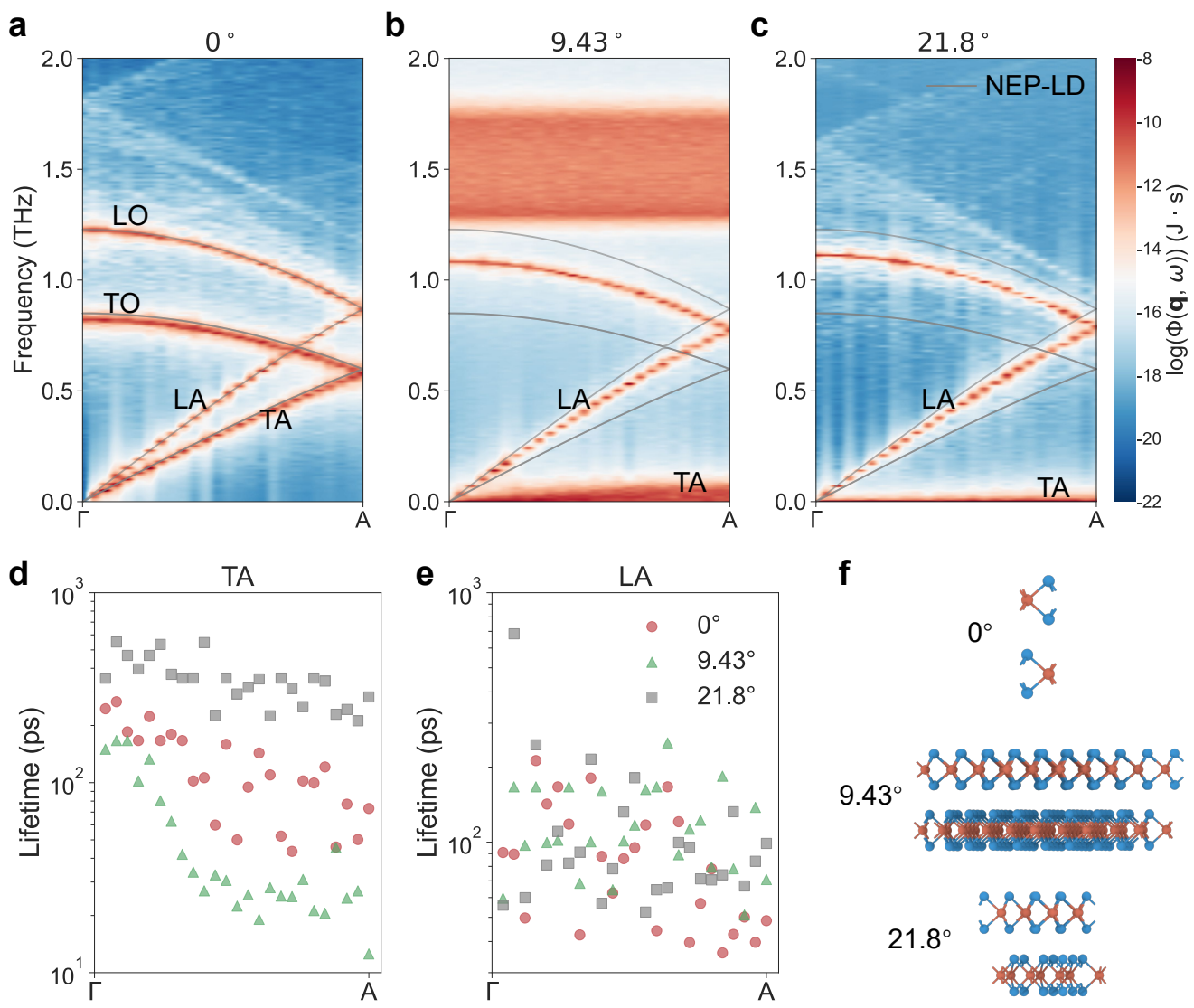

FIG. 5. The effect of the twist angle on the phonon SED and lifetimes of bilayer $MoS_2$ along the Γ to A direction at 300 K. For clarity, the coupling strength $\eta$ here is 1.0. Panels **a**–**c** correspond to twist angles of 0°, 9.43°, and 21.8°, respectively. The gray solid line represents the results of lattice dynamics calculations. **d**–**e** Lifetimes of the TA and LA phonon modes at different twist angles. **f** Unit cells corresponding to different twist angles for bilayer $MoS_2$.

Concurrently, in Figure 5e, the phonon lifetimes of LA modes remain within the same order of magnitude across three twist angles, similar to the previous work [6, 70]. Figure 5f displays the unit cells employed in SED calculations, with the $\theta = 9.43°$ configuration containing 222 atoms. This systematic investigation confirms the capability of PYSED for probing twist-angle-dependent thermal transport mechanisms in layered materials.

## D. MOFs

MOFs represent a class of materials distinguished by high porosity, strong anharmonicity, and intrinsically low thermal conductivity [7]. Their complex unit cells render the thermal conductivity inadequately described by the Boltzmann transport equation (BTE) under the quasiparticle picture, typically requiring simultaneous consideration of particle transport and coherent interband contributions [73]. In disordered solids and strongly anharmonic crystals, when phonon interband spacings become smaller than the linewidths, interband tunneling occurs between phonon branches with different frequencies due to the wave nature of phonons. In this scenario, particlelike and wavelike mechanisms coexist and are both relevant, with the contribution of interband coherence to thermal conductivity being non-negligible [73–75]. Here, we employ the SED method to present the phonon dispersion of two MOFs (MOF-5 and HKUST-1) at different temperatures and investigate whether their corresponding phonon lifetimes can reveal a picture of the coexistence of particlelike and wavelike transport.

Figure 6a–b illustrates the SEDs of MOF-5 and HKUST-1 at 300 K, which align closely with the 0 K lattice dynamics predictions, particularly for the low-frequency acoustic branches. As the temperature rises to 500 K, the high-frequency optical branches exhibit significant broadening, indicating strong anharmonicity at high temperatures. Moreover, as shown in Figure 6c–d, the phonon lifetimes derived from Lorentzian fitting decrease with increasing temperature, especially within the sub-0.5 THz low-frequency regime. According to the "Wigner limit in time" criterion, the region well above the red dashed line represents phonons that propagate particlelike and mainly contribute to the population's conductivity, while those below the line represent tunneling wavelike phonons responsible for coherent conductivity. The gray solid line, $\tau = 1/\omega$, marks the center of the nonsharp Ioffe-Regel limit in time. Regions above this line correspond to well-defined phonons, while areas below it represent the so-called overdamped regime, where phonons are no longer well-defined quasi-particle excitations [73, 75].

In the thermal transport regime diagram of MOFs (see Figure 6c–d), phonon modes below 0.5 THz are predominantly acoustic and exhibit particlelike behavior, whereas supra-0.5 THz modes are dominated by optical phonons with certain wave-like tunneling features. As temperature increases, the population of particlelike modes decreases, whereas wavelike modes grow, consistent with previous works [73, 75, 76]. Additionally, all phonon lifetimes in MOFs reside above the Ioffe-Regel limit threshold, demonstrating the applicability of the Wigner transport equation [75] or quasi-harmonic Green-Kubo approach [74] for probing the thermal transport properties of MOFs. These results validate that the SED method effectively establishes the thermal transport regime diagram for MOFs, which can be analogously extended to qualitative analysis of strongly anharmonic materials such as perovskites [73], solid-state electrolytes [77], crystalline organic semiconductors [76], skutterudites [78], and zeolite frameworks [79].

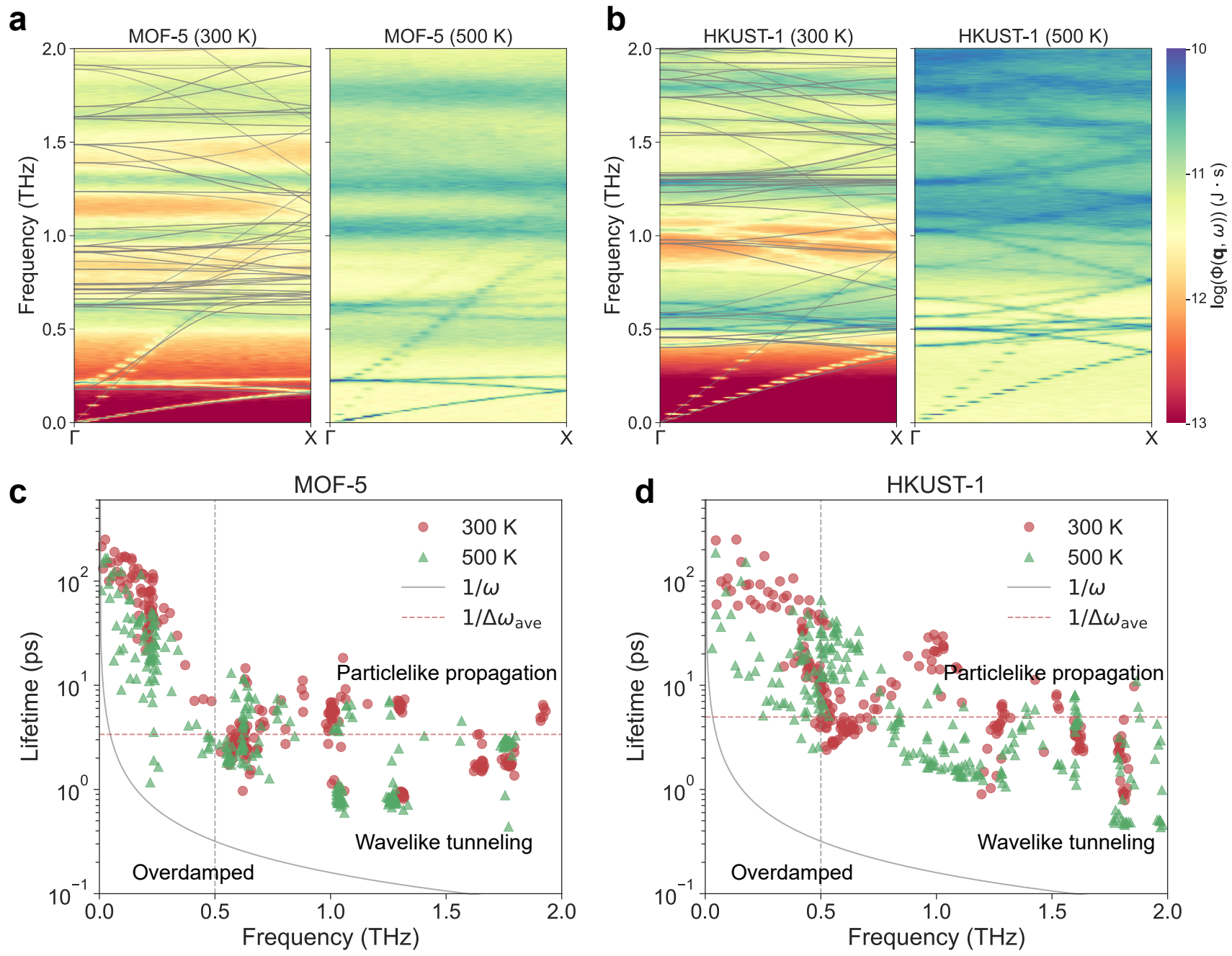

FIG. 6. **a**–**b** Phonon SED of MOF-5 and HKUST-1 at different temperatures. The solid gray line represents the lattice-dynamics calculation results. **c**–**d** Phonon lifetimes for MOF-5 and HKUST-1 at various temperatures derived from Lorentzian fitting. The gray dashed line at 0.5 THz serves as a visual guide to distinguish low-frequency and long-wavelength phonons. The horizontal red dashed line is the Wigner limit in time $\tau = 1/\Delta\omega_{\rm av}$, where $\Delta\omega_{\rm av} = \frac{\omega_{\rm max}}{3N_{\rm at}}$. Here, $\Delta\omega_{\rm av}$ is the inverse average interband spacing, representing the center of the nonsharp particle-wave crossover for phonons. The $\omega_{\rm max}$ and $N_{\rm at}$ represent the maximum phonon frequency and the number of atoms in the primitive cell, respectively. The gray solid line, $\tau = 1/\omega$, marks the center of the nonsharp Ioffe-Regel limit in time [73] with $\omega$ being the angular frequency.

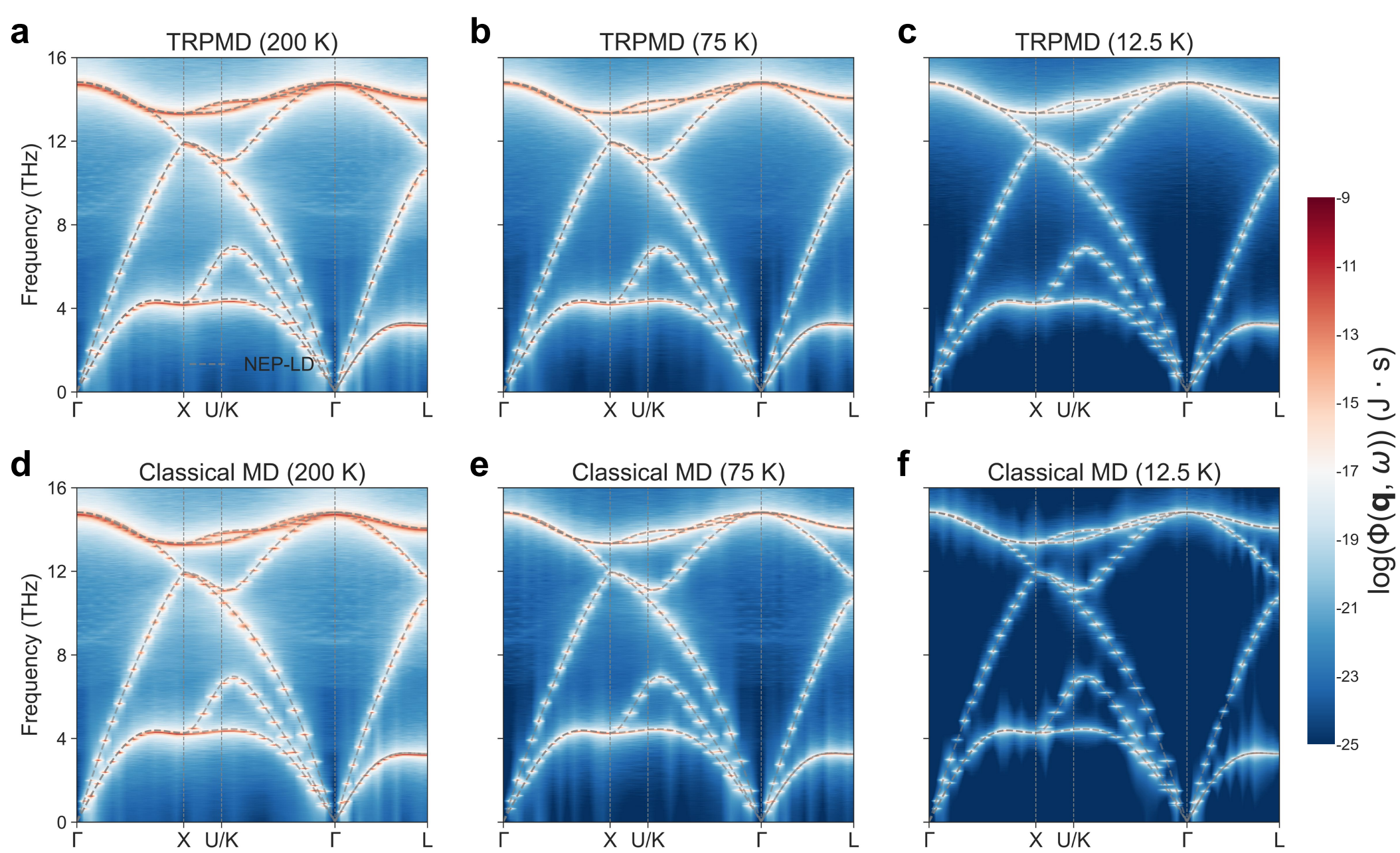

FIG. 7. **a**–**c** Phonon SED at different temperatures via TRPMD simulations. **d**–**f** Classical MD calculated phonon SED under varying temperatures.

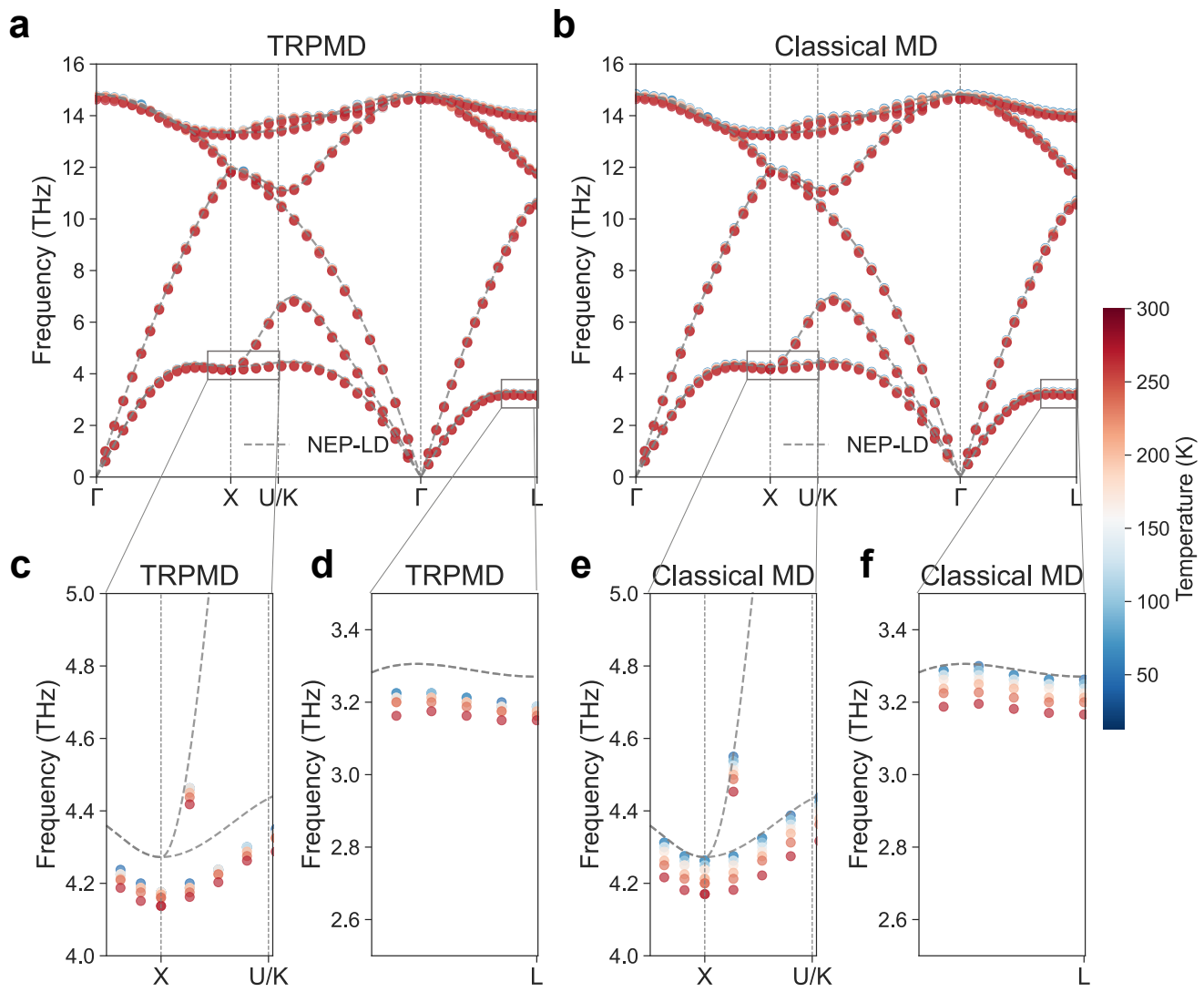

FIG. 8. A comparison between the SED-derived phonon dispersion and lattice-dynamics calculated results at varying temperatures. Panels **a** and **b** present the **q**-point-resolved frequencies obtained by applying Lorentzian fitting to the SED results from TRPMD and classical MD simulations, respectively, covering eight temperature points (300, 200, 150, 100, 75, 50, 25, and 12.5 K). **c**–**d** Magnifications of the TA mode near the reciprocal points $X$ and $L$ from TRPMD. **e**–**f** Magnifications of the acoustic modes near the reciprocal points $X$ and $L$ from classical MD.

## E. Silicon

Silicon stands as the cornerstone semiconductor material, underpinning the development of modern semiconductor devices and integrated circuit technologies. Although its relatively large atomic mass has led to the common assumption that NQEs in silicon are insignificant, the reality is far more complex [80, 81]. Here, we employ SED analysis driven by quantum and classical trajectories to investigate the phonon properties of silicon at low temperatures. For a more comprehensive discussion of the PIMD and TRPMD simulations used in this section to account for NQEs, readers are encouraged to consult Ref. [58].

Figure 7 compares quantum/classical phonon SED and lattice dynamics results (0 K) at various temperatures, demonstrating good agreement and validating the broad applicability of the SED approach. In particular, both TRPMD and classical SED exhibit minor frequency softening relative to zero-temperature frequencies at elevated temperatures, particularly in high-frequency regimes, attributable to the anharmonicity of temperatures. At ultralow temperatures (12.5 K), however, the TRPMD-SED maintains these frequency shifts while classical MD results nearly approach the zero-temperature baseline. For further clarity, we performed Lorentzian fitting to the SED spectra at each **q** point. The extracted temperature-dependent phonon dispersions, presented in (Figure 8a–b), highlight the excellent frequency-fitting accuracy achieved by PYSED package. Figure 8c–f display magnified views of the acoustic modes at the $X$ and $L$ points, obtained from TRPMD and classical MD, where $X$ and $L$ represent reciprocal points (1/2, 0, 1/2) and (1/2, 1/2, 1/2), respectively. While both methods predict increasing phonon frequencies (hardening) with decreasing temperature, classical MD-SED frequencies asymptotically converge to the lattice dynamics baseline as $T \to 0\,\mathrm{K}$, whereas TRPMD-SED retains substantial offsets due to the quantum fluctuations.

To establish quantitative insights, we calculated temperature-dependent phonon frequency shifts for the TA and LA modes at the $X$ and $L$ points. The frequency shifts were determined by subtracting the zero-temperature frequency from the frequencies fitted at the respective temperature. In Figure 9a–b, as $T \to 0\,\mathrm{K}$, the phonon frequency shifts for the four typical modes ($X$-TA, $X$-LA, $L$-TA, and $L$-LA) from classical MD-SED tend toward zero. In the classical case, there is no zero-point energy, and atomic displacements decrease toward zero as the temperature approaches absolute zero. However, in the quantum scenario, the frequency shift does not reach zero due to quantum fluctuations, and these shifts tend to stabilize at very low temperatures. This expected behavior is observed in the TRPMD-SED results (Figure 9a–b), demonstrating its accurate capture of NQEs. The temperature-dependent relative phonon shifts for the TA modes at the $X$ and $L$ points were further calculated using $(\epsilon_T - \epsilon_{100K})/\epsilon_{100K}$, where $\epsilon_T$ denotes the phonon frequency measured at temperature $T$, and $\epsilon_{100K}$ represents the corresponding value at a reference temperature of 100 K. This relative change (see Figure 9c–d) agrees well with the results from inelastic neutron scattering (INS) measurements [80]. Additionally, as the temperature increases, the relative phonon frequency shifts become negative, indicating phonon softening and incorporating the anharmonicity of temperatures.

Apart from phonon frequency shifts, we also extracted temperature-dependent phonon lifetimes (see Figure 9e–g) through Lorentzian fitting. For comparison, we calculated the quantum and classical lifetimes at different temperatures within the ALD-BTE framework using the KALDO package [82], with quantum and classical statistics based on Bose-Einstein and Boltzmann distributions, respectively. The BTE calculations were performed using a conventional silicon unit cell (8 atoms) with $6 \times 6 \times 6$ supercells and a $15 \times 15 \times 15$ q-point grid. At 200 K, the SED-derived phonon lifetimes demonstrate comparable magnitudes to the BTE-calculated ones (see Figure 9e and Figure S5a), while in the low-frequency regime ($< 5\,\mathrm{THz}$), SED systematically yields shorter lifetimes than BTE predictions due to its inherent inclusion of full-order scattering events. As demonstrated in Figure 9e–g, as the temperature approaches zero, the lifetimes obtained from classical MD-SED gradually increase and tend to diverge (Figure 9g). In contrast, due to the consideration of NQEs, the phonon lifetimes in TRPMD-SED remain finite as the temperature approaches zero, and the lifetimes for higher frequencies gradually decrease. The phonon lifetimes predicted by the ALD-BTE framework also increase and tend to diverge as the temperature approaches zero (see Figure S5). This is likely due to the lack of high-order anharmonic terms, as KALDO only includes third-order force constants and does not account for boundary scattering. The results presented above demonstrate that the SED method effectively captures both quantum and classical dynamical properties, depending on the nature of the input trajectories.

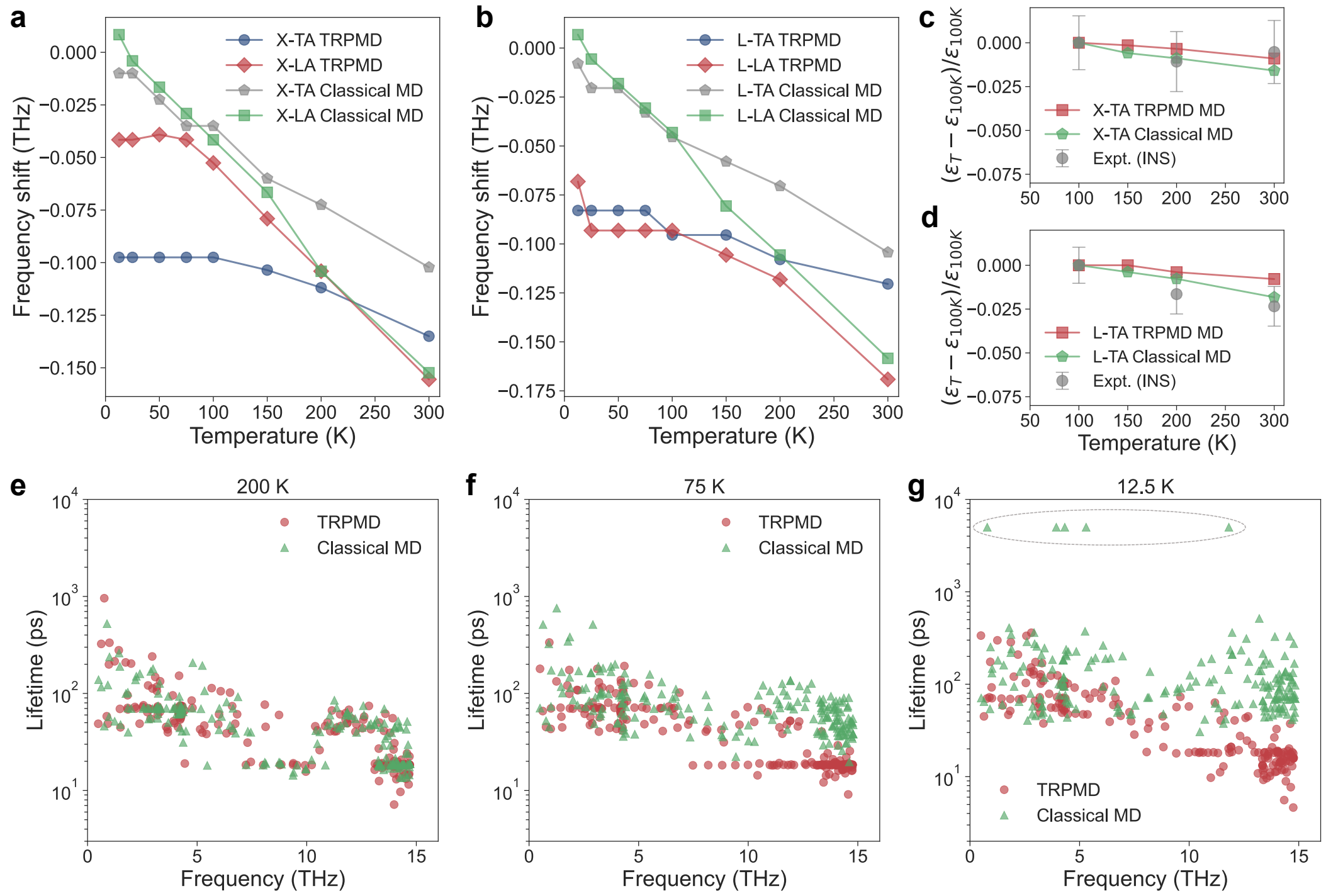

FIG. 9. **a–b** Phonon frequency shifts of the TA and LA modes at the $X$ and $L$ high-symmetry points in bulk silicon. **c–d** Temperature-dependent relative phonon frequency shifts $(\epsilon_T - \epsilon_{100K})/\epsilon_{100K}$ of the TA modes at the $X$ and $L$ high-symmetry points. The experimental data were obtained through inelastic neutron scattering measurements [80]. **e–g** Phonon lifetimes at different temperatures calculated from TRPMD and classical MD driven SED.

## V. CONCLUSIONS

In summary, we present PYSED, a Python-based package built upon the SED method, designed to analyze specific phonon-mode information from large-scale MD trajectories, enabling convenient calculation of kinetic-energy-weighted phonon dispersion and derivation of phonon lifetime. Leveraging highly efficient machine-learned NEP models to drive MD simulations, we validate the generality of the SED approach and the efficacy of the PYSED package by investigating phonon characteristics across diverse systems, including 1D CNTs, 2D graphene, $h$-BN, and $MoS_2$, as well as 3D MOFs and silicon.

Using NEP-driven MD simulations, the SED method reveals that compressive strain reduces phonon lifetimes in CNTs due to intensified scattering induced by localized buckling. In 2D systems, it yields phonon dispersions for graphene and $h$-BN that closely match lattice dynamics calculations and experimental measurements, while capturing the pronounced collapse of phonon modes in $MoS_2$ under varying interlayer coupling and twist angles. The SED method also effectively establishes the thermal transport spectrum of MOFs, distinguishing particlelike and wavelike propagation regions in fitted phonon lifetimes. Leveraging efficient NEP-PIMD simulations within the GPUMD package [23], we demonstrate that SED, depending on the properties of input trajectories, accurately captures both quantum and classical dynamics in the phonon characteristics of bulk silicon.

In this work, starting from the theoretical derivation of SED, we implemented the PYSED package, providing an accurate and scalable tool for studying the phonon-mode-level nature of various materials from MD trajectories generated by MLPs. Currently, the SED method implemented in PYSED is limited to periodic systems with long-range order and definable unit cells. For amorphous systems, liquids, and the like, the dynamic structure factor approach implemented in DYNASOR package [18, 19] can be consulted, as it is suitable for all systems. Notably, during the preparation of our manuscript, the PYSED package has already been applied to extract phonon-mode information from molecular dynamics (MD) trajectories in various systems [7, 48, 49, 55, 67, 83–88].

## VI. SUPPLEMENTARY MATERIAL

In the Supplementary Material, we provide additional figures that complement and support the findings presented in the main text. Specifically, the Supplementary Material contains the following items:

1. **Figure S1**: The training parity plot of the NEP model for $MoS_2$ system.
2. **Figure S2**: Comparison of the CNTs' phonon dispersion from different supercells with the SED results.
3. **Figure S3**: Lorentzian fitting of the SED peaks for CNT systems.
4. **Figure S4**: Lorentzian fitting for $MoS_2$ systems with different twist angles at various **q** points.
5. **Figure S5**: Temperature-dependent phonon lifetimes from the Boltzmann transport equation framework.

## ACKNOWLEDGMENTS

T.L. sincerely thanks Dr. Zekun Chen for the discussions and support regarding the use of KALDO [82]. T.L. sincerely thanks Dr. Tyler C. Sterling for his invaluable help with the development of the PYSED code. T.L. also expresses gratitude to Dr. Penghua Ying for valuable discussions on the MOFs examples. T.L., K.X., and J.X. acknowledge support from the National Key R&D Project from the Ministry of Science and Technology of China (No. 2022YFA1203100), RGC GRF (No. 14220022), RGC Fund for Joint Research Labs (JLFS/E-402/24), and the CUHK PhD Studentship and Postdoctoral Fellowship. W.O. acknowledges the support from the National Natural Science Foundation of China (Nos. 12472099 and U2441207), the Fundamental Research Funds for the Central Universities (Nos. 2042025kf0013 and 600460100), and the start-up fund of Wuhan University. Some computations were conducted at the Supercomputing Center of Wuhan University, the National Supercomputer TianHe-1(A) Center in Tianjin, and the Computing Center in Xi'an.

### Code Availability

The source code and documentation for PYSED are available at https://github.com/Tingliangstu/pySED and https://pysed.readthedocs.io, respectively.

## Supplementary Material:

# PYSED: A tool for extracting kinetic-energy-weighted phonon dispersion and lifetime from molecular dynamics simulations

Ting Liang *[1], Wenwu Jiang *[2], Ke Xu[1], Hekai Bu[2], Zheyong Fan[3], Wengen Ouyang †[2, 4], and Jianbin Xu ‡[1]

[1] *Department of Electronic Engineering and Materials Science and Technology Research Center, The Chinese University of Hong Kong, Shatin, N.T., Hong Kong SAR, 999077, P. R. China*
[2] *Department of Engineering Mechanics, School of Civil Engineering, Wuhan University, Wuhan, Hubei 430072, China*
[3] *College of Physical Science and Technology, Bohai University, Jinzhou 121013, P. R. China*
[4] *State Key Laboratory of Water Resources Engineering and Management, Wuhan University, Wuhan, Hubei 430072, China*

# Contents

*These authors contributed equally to this work.
†Email: w.g.ouyang@whu.edu.cn
‡Email: jbxu@ee.cuhk.edu.hk

# Supplemental Figures

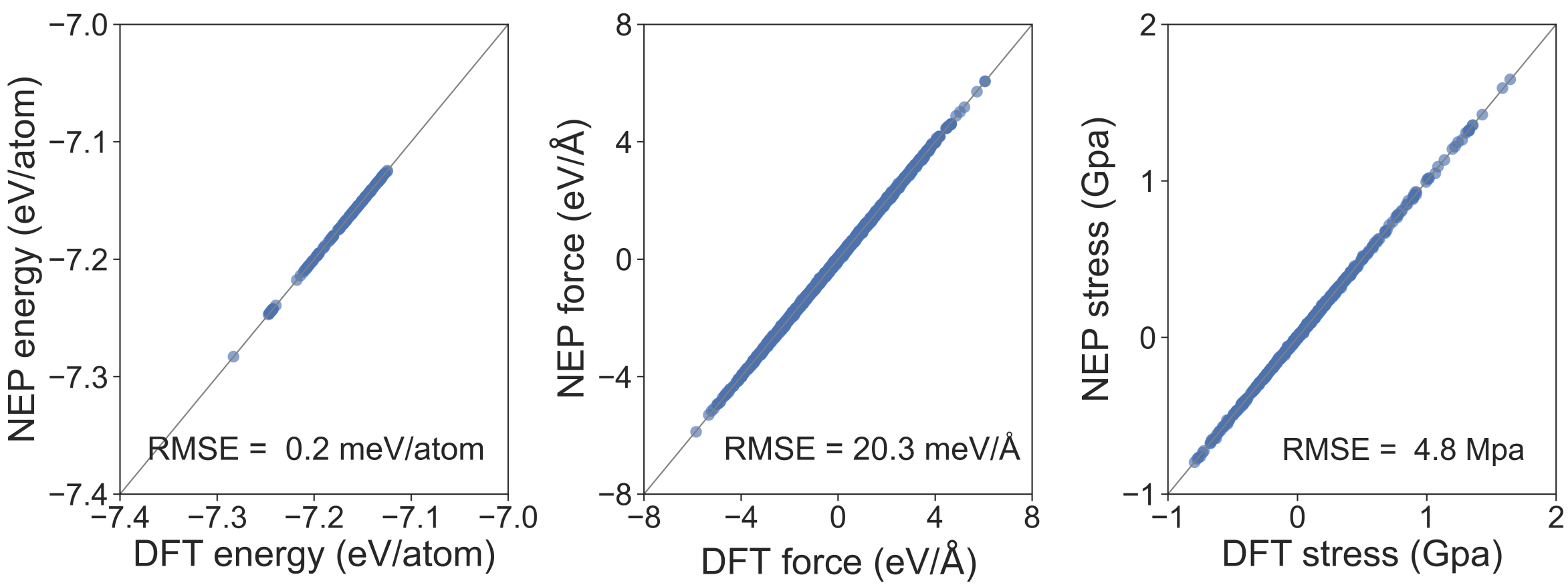

Figure S1: A comparison of the energy, force, and stress predicted by the NEP model for the $MoS_2$ system with the DFT reference data. In each panel, RMSE denotes the root mean square error.

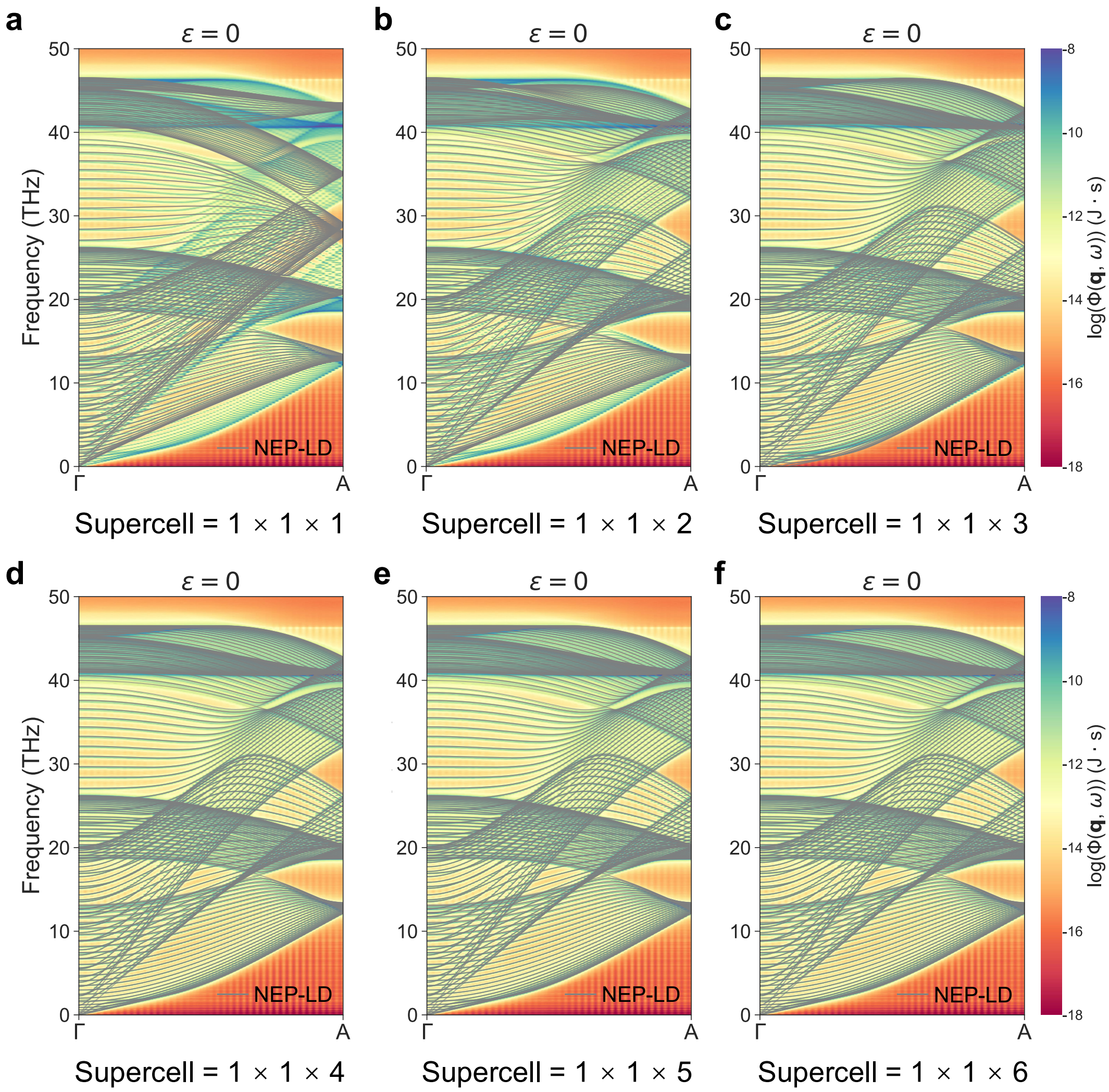

Figure S2: **a–f** Comparison of the phonon dispersion of CNTs at zero strain, obtained from lattice dynamics calculations using different supercells (labeled as "NEP-LD"), with SED calculations (300 K). The supercell sizes of CNTs used in the lattice dynamics calculations are specified below each subplot. From panels **a** to **f**, as the supercell size used in lattice dynamics calculations reaches or exceeds $1 \times 1 \times 4$ , the CNTs' phonon dispersion at 0 K (gray solid line), matches closely with the SED calculations. Using an insufficiently large supercell in lattice dynamics calculations results in a truncated force constant, preventing adequate sampling of long-range interatomic interactions in real space. Therefore, careful validation of the supercell sizes in lattice dynamics calculations is necessary to obtain phonon dispersion results comparable to those from SED.

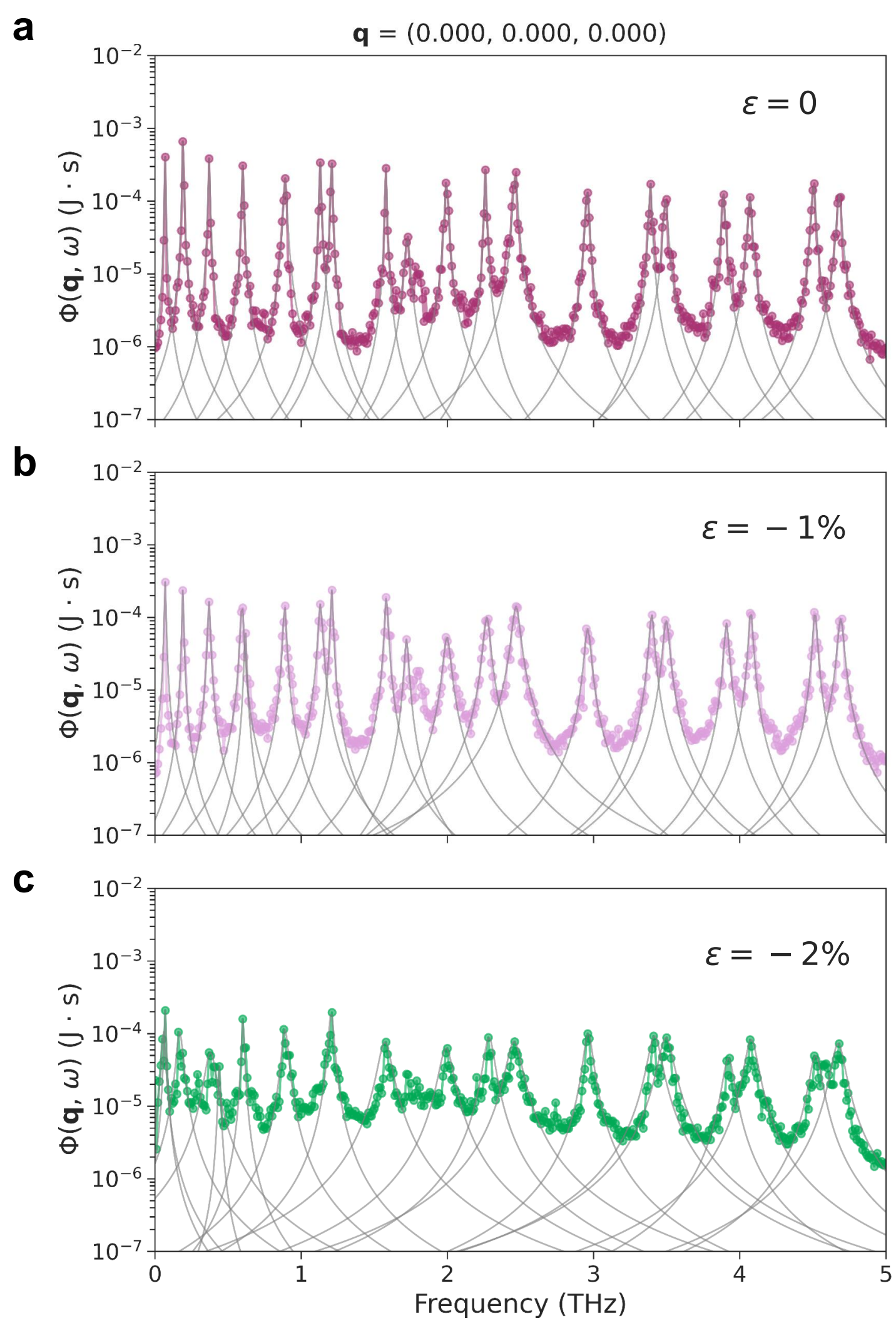

Figure S3: **a–c** Lorentzian fitting of the SED peaks at a typical $\boldsymbol{q}$ point (0, 0, 0) for CNT systems under different compressive strains. The gray solid line represents the Lorentzian fitting of the SED peaks. With increasing compressive strain, the broadening of SED peaks becomes more pronounced, particularly in the relatively higher-frequency regions.

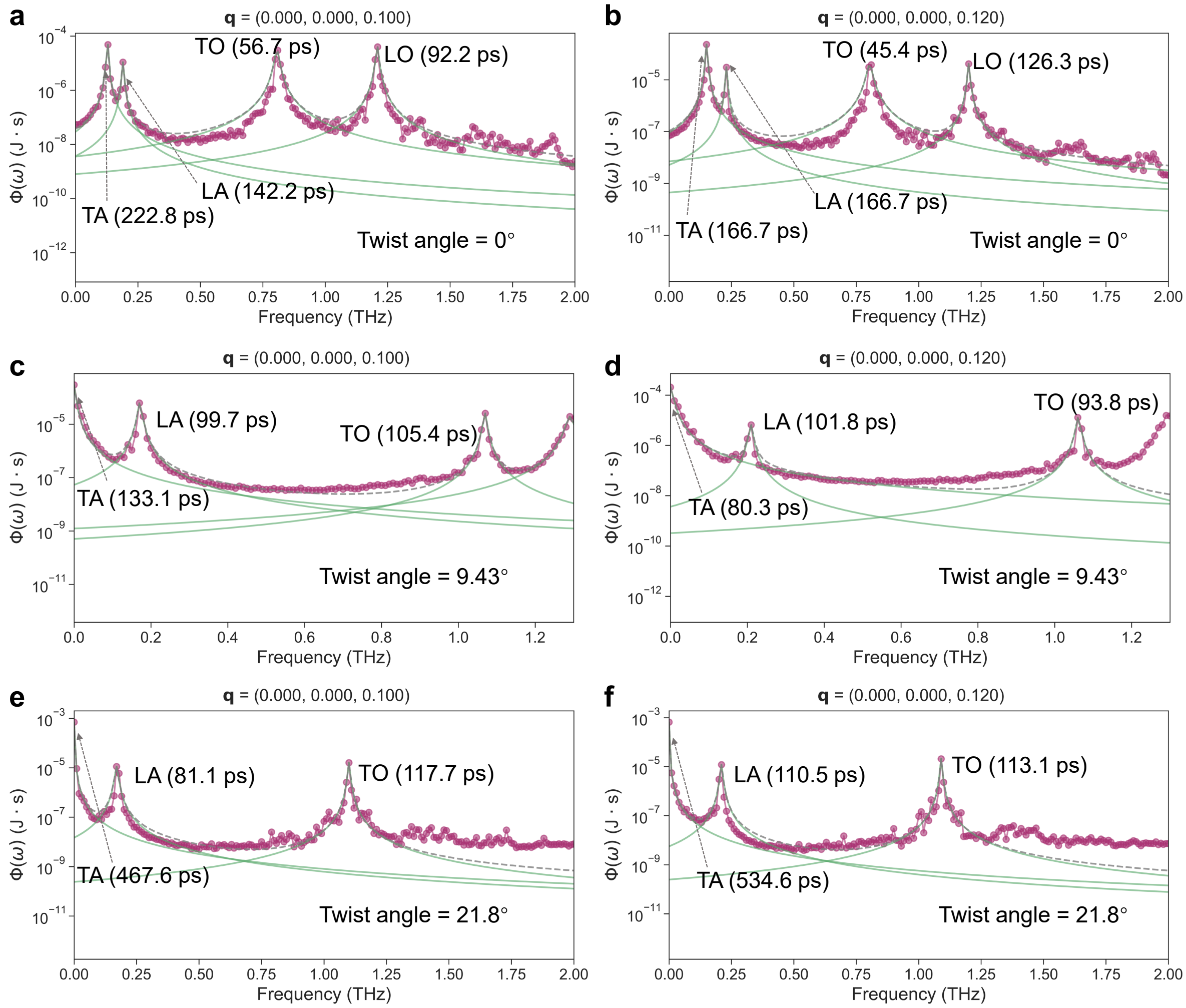

Figure S4: **a–f** Lorentzian fitting for bilayer $MoS_2$ systems with different twist angles at various **q** points. The lifetimes corresponding to different phonon branches are marked. The green solid line represents the Lorentzian fitting of a single SED peak, while the grey dashed line is the superposition of Lorentzian fittings for all peaks. With the built-in module of the PYSED package, the SED peaks can be automatically identified and fitted accurately.

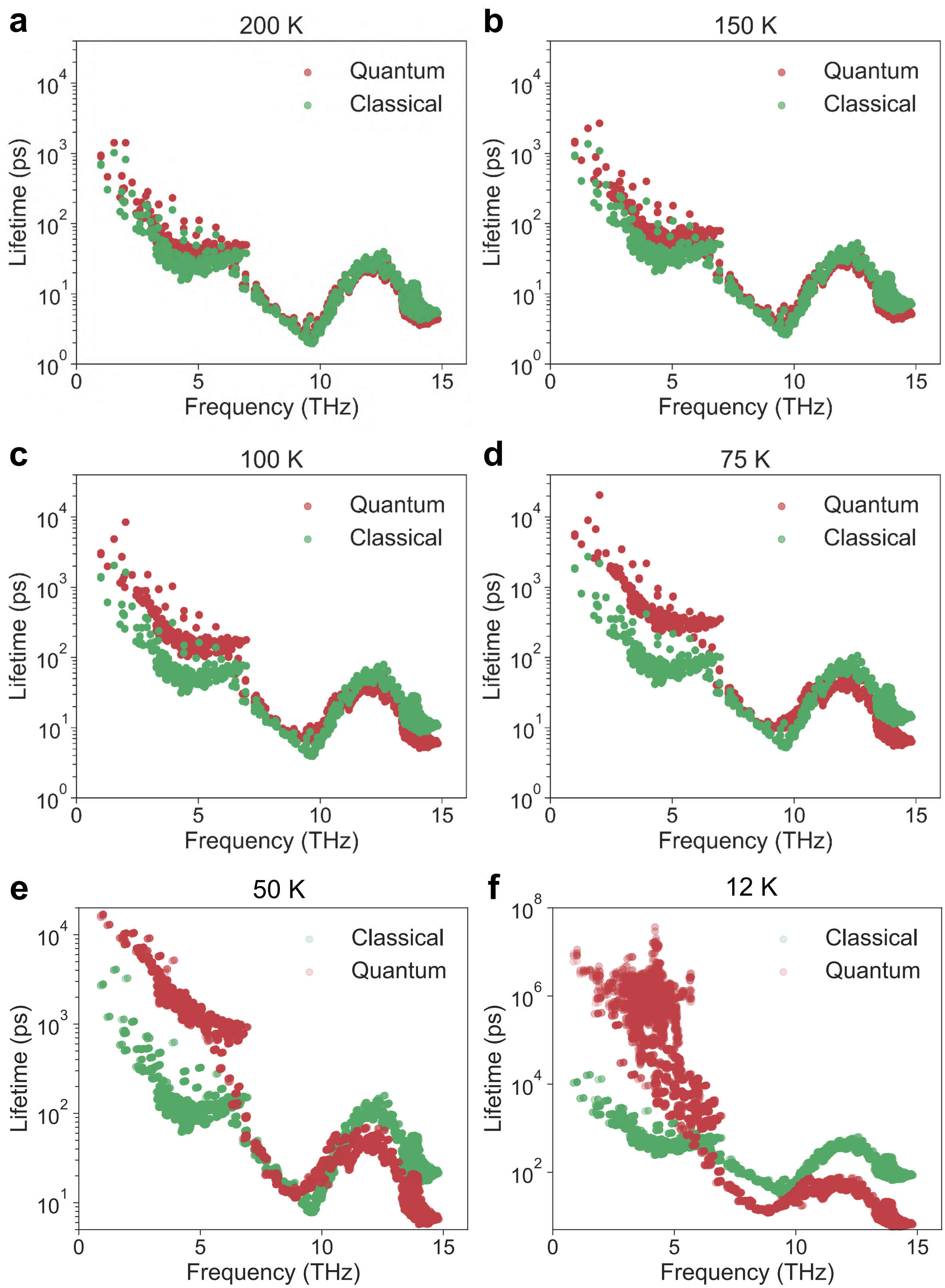

Figure S5: **a–f** Temperature-dependent phonon lifetimes with quantum and classical statistics computed under the Boltzmann transport equation framework. In the KALDO package, both quantum and classical phonon lifetimes are provided based on the relaxation time approximation method.